\documentclass[fleqn,usenatbib]{mnras}

\usepackage{newtxtext,newtxmath}

\usepackage[T1]{fontenc}

\DeclareRobustCommand{\VAN}[3]{#2}
\let\VANthebibliography\thebibliography
\def\thebibliography{\DeclareRobustCommand{\VAN}[3]{##3}\VANthebibliography}

\usepackage{graphicx}	
\usepackage{amsmath}	
\hypersetup{hypertexnames=false}

\title[State-dependent X-ray timing in NGC 1566]{State-dependent broadband X-ray timing reconfiguration in the changing-look AGN NGC 1566}

\author[Y. Tao et al.]{%
Y. Tao,$^{1}$
J. Tang,$^{1}$\thanks{E-mail: tj168@163.com},
X. Wei,$^{1}$
and X.-H. Zhang$^{1}$%
\\
$^{1}$School of Physics and Telecommunication Engineering, Shaanxi University of Technology, Hanzhong 723000, People's Republic of China
}

\date{Accepted 2026 June 11. Received 2026 June 10; in original form 2026 March 16}

\pubyear{\the\year{}}

\begin{document}
\label{firstpage}
\pagerange{\pageref{firstpage}--\pageref{lastpage}}
\maketitle

\begin{abstract}
NGC 1566 has shown dramatic X-ray spectral changes during its recent changing-look outburst, but the evolution of its broadband X-ray timing properties remains poorly constrained. We combine long-term \textit{Swift}/XRT monitoring with high-time-resolution \textit{XMM-Newton} observations to construct one pre-outburst Dim broadband PSD and two outburst broadband reconstructions associated with the O1 peak and O2 decay observations. In the outburst reconstructions, the same \textit{Swift}/XRT up-state monitoring segment provides the low-frequency constraint, while the O1 and O2 \textit{XMM-Newton} observations provide phase-specific high-frequency constraints. Using PSRESP forward modelling with the observed sampling windows, we test bending-power-law PSD models in the soft (0.3--2 keV) and hard (2--10 keV) bands. The Dim and O2-associated reconstructions are acceptably described by bending-power-law solutions, whereas the O1 peak observation does not yield a robust bend-frequency measurement. For the accepted Dim and O2-associated solutions, the preferred bend frequency shifts from \(\sim2.0\times10^{-5}\) to \(\sim2.7\times10^{-7}\,\mathrm{Hz}\) in the soft band, and from \(\sim2.1\times10^{-5}\) to \(\sim2.7\times10^{-7}\,\mathrm{Hz}\) in the hard band, implying a substantially longer characteristic variability timescale in the O2-associated reconstruction. This consistent shift in both energy bands suggests that the timing evolution is not confined to the soft-excess component alone, but reflects a broader change in the X-ray variability structure. Together with previous spectral studies, these results point to a transient reconfiguration of the disc--corona variability timescale during the changing-look transition in NGC 1566.
\end{abstract}




\section{Introduction}

Changing-look active galactic nuclei (CL AGNs) provide a valuable opportunity to probe rapid changes in the inner accretion flow across different radiative states. 
They are characterized by the appearance or disappearance of broad optical emission lines, often accompanied by large continuum changes on timescales much shorter than expected from standard thin-disc viscous evolution. 
The discovery of the changing-look quasar SDSS J015957.64+003310.5 demonstrated that quasar-like objects can undergo such transitions on humanly accessible timescales \citep{LaMassa2015}, while subsequent statistical studies have shown that large-amplitude optical variability is closely connected to the CLAGN phenomenon \citep{MacLeod2016}. 
In many cases, multiwavelength observations favour intrinsic accretion-flow changes over line-of-sight obscuration alone as the primary driver of the transition (e.g. Ruan et al. 2016; Noda \& Done 2018; Ricci \& Trakhtenbrot 2023). Recent long-term multiwavelength monitoring of NGC 3822 similarly associated the changing-look transitions with changes in the Eddington ratio rather than with variable obscuration \citep{Layek2025}.

NGC 1566 is one of the best-studied nearby changing-look Seyfert galaxies, particularly following its extreme 2018 outburst \citep{Oknyansky2019,Parker2019}. 
Broadband X-ray studies showed that, near the peak of the outburst, the source displayed the spectral properties of a Seyfert 1 nucleus and that its behaviour was most likely associated with intrinsic accretion-flow evolution \citep{Parker2019}. 
Later work further showed substantial changes in the disc, soft excess, and broadband continuum, together with systematic evolution of the warm and hot Comptonizing regions, supporting a reconfiguration of the inner radiative structure \citep{Jana2021,TripathiDewangan2022}. 
These spectral results indicate that the 2018 event was not a simple luminosity rescaling, but involved substantial changes in the disc--corona system.

In contrast, the broadband X-ray timing behaviour of NGC 1566 remains much less well constrained. 
X-ray variability in accreting black holes is commonly interpreted in terms of stochastic fluctuations generated over a range of disc radii and propagating inward through the accretion flow \citep{Lyubarskii1997,ArevaloUttley2006,Uttley2005}. 
In this framework, the power spectral density (PSD) encodes characteristic variability timescales of the inner accretion flow. 
In particular, the PSD bend frequency $f_{\rm b}$ defines a characteristic timescale, $t_{\rm b}=1/f_{\rm b}$, and is empirically linked to black-hole mass and accretion rate in AGN and X-ray binaries \citep{McHardy2004,McHardy2006}. 
A state-dependent shift in $f_{\rm b}$ therefore implies a change in the dominant X-ray variability timescale, which may reflect changes in the effective size of the variable emitting region, the propagation path of accretion-rate fluctuations, or the coupling between the disc and corona. 
Energy-dependent PSD measurements can further test whether the soft and hard X-ray variability components respond coherently or trace distinct regions of the accretion flow \citep[e.g.][]{ArevaloUttley2006,Uttley2014}.

NGC 1566 is therefore an especially useful target for testing whether a changing-look transition is accompanied by a measurable reconfiguration of broadband X-ray timing properties. 
Here we combine long-term \textit{Swift}/XRT monitoring with high-time-resolution \textit{XMM-Newton} observations to investigate the broadband X-ray PSD of NGC 1566 in three representative epochs: a pre-outburst Dim phase, the O1 peak observation, and the O2 decay observation. The Dim broadband PSD is constructed from the pre-outburst \textit{Swift}/XRT interval and the Dim \textit{XMM-Newton} observation. For the outburst, the available long-baseline \textit{Swift}/XRT monitoring covers the overall bright episode and early decline; it is therefore used as a common low-frequency constraint for two separate broadband reconstructions, one associated with the O1 peak \textit{XMM-Newton} observation and one associated with the O2 decay \textit{XMM-Newton} observation. These reconstructions should not be interpreted as independent O1-only and O2-only low-frequency PSD measurements, but as tests of whether the phase-specific high-frequency variability can be connected to the common outburst-timescale variability within a single stationary PSD parametrization. This approach allows us to examine whether the spectral reconfiguration already established in NGC 1566 is accompanied by a corresponding change in the characteristic X-ray variability timescale, while keeping explicit the limitations imposed by the available sampling.
\section{Data and methods}

\subsection{Data reduction and state selection}

We combine long-term \textit{Swift}/XRT monitoring with high-time-resolution \textit{XMM-Newton} observations to characterize the broadband X-ray variability of NGC 1566. 
The timing intervals are selected to trace the long-term X-ray evolution of the source around its 2018 changing-look event. 
We define MJD 56185--57233 as the pre-outburst Dim interval, corresponding to the low-flux state before the 2018 event. 
The 2018 outburst was identified from the optical, UV, and X-ray evolution of NGC 1566 \citep{Oknyansky2019}, and the source reached a strong X-ray brightening in 2018 June, with the X-ray intensity increasing by a factor of $\sim25$--30 relative to the quiescent level \citep{Jana2021}. 
We therefore define MJD 58293--58437 as the up-state monitoring interval, covering the main bright episode and the early post-maximum decline. 
The post-maximum evolution further shows substantial variability and later re-brightenings \citep{Oknyansky2020}, while broadband X-ray spectral studies distinguish the June peak-outburst epoch from the subsequent declining phase \citep{TripathiDewangan2022}. 
Accordingly, the full MJD 58293--58437 interval is used as the \textit{Swift}/XRT up-state segment for the low-frequency PSD analysis, whereas the two high-time-resolution \textit{XMM-Newton} observations within this interval are kept separate and are used to represent the O1 peak and O2 decay phases at high frequencies.

The \textit{Swift}/XRT monitoring light curves are obtained from the online products of \citet{Evans2009}. 
For each selected interval, we generate background-subtracted and exposure-corrected count-rate light curves in the soft (0.3--2 keV) and hard (2--10 keV) bands, matching the energy bands used in the PSD analysis. 
These \textit{Swift}/XRT light curves provide the low-frequency PSD constraints. 
Because the \textit{Swift}/XRT data are irregularly sampled, their sampling window is explicitly included in the PSRESP forward modelling. 
The up-state interval is sampled with typical separations of several days to about one week, whereas the longer Dim interval contains larger temporal gaps. 
Individual \textit{Swift}/XRT visits have typical effective exposures of order $10^3$ s. 
In the PSRESP simulations described below, the synthetic light curves are sampled at the actual \textit{Swift}/XRT observing times, preserving both the cadence and the long gaps of the monitoring data.

For the \textit{XMM-Newton} data, we reduce the EPIC-pn observations associated with the three representative phases: ObsID 0763500201 for the Dim phase, ObsID 0800840201 for the O1 peak phase, and ObsID 0820530401 for the O2 decay phase. 
The data are reduced using SAS v22.1.0, following standard procedures including the generation of calibration files, event reconstruction, flare filtering based on the 10--12 keV background light curve, and event selection with \texttt{PATTERN <= 4} and \texttt{FLAG = 0}. 
Pile-up is assessed using \texttt{epatplot}, with particular attention to the central part of the 36 arcsec source extraction region, and no significant pile-up requiring an annular extraction region is found. 
We therefore extract the final soft and hard light curves from a circular source region with a radius of 36 arcsec and a nearby source-free background region, and correct them using \texttt{epiclccorr} for background subtraction and instrumental effects. 
The resulting \textit{XMM-Newton} light curves are binned at 50 s and used to constrain the high-frequency PSD.

\begin{figure*}
    \centering
    \includegraphics[width=\textwidth]{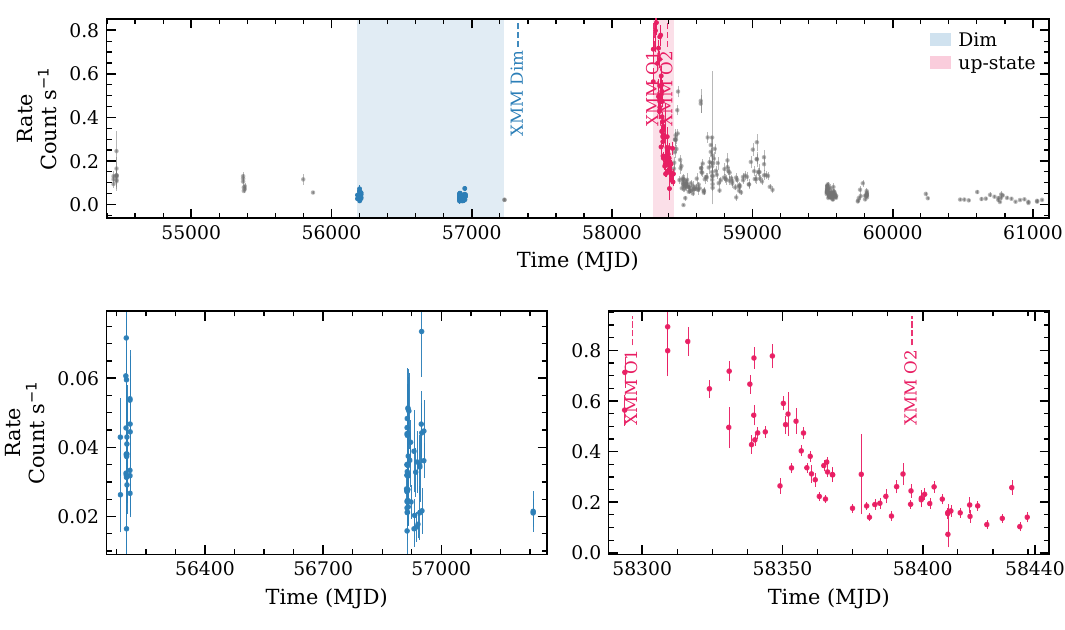}
    \caption{\textit{Swift}/XRT 2--10 keV light curve of NGC 1566. The blue shaded interval marks the pre-outburst Dim monitoring segment used as the low-frequency constraint for the Dim reconstruction. The pink shaded interval marks the outburst up-state monitoring segment used as the common low-frequency constraint for the O1- and O2-associated reconstructions. The lower panels zoom in on the selected Dim and up-state intervals. Short markers indicate the epochs of the \textit{XMM-Newton} observations; the Dim \textit{XMM-Newton} epoch lies outside the selected \textit{Swift}/XRT Dim monitoring segment and is therefore not shown in the Dim zoom panel.}
    \label{fig:swift_lc}
\end{figure*}

The Dim \textit{XMM-Newton} observation, ObsID 0763500201, was obtained on 2015 November 5 (MJD 57331), about 98 d after the end of the selected \textit{Swift}/XRT Dim monitoring segment (MJD 56185--57233). Together, the long-baseline \textit{Swift}/XRT monitoring and the short, continuous \textit{XMM-Newton} observations provide complementary low- and high-frequency coverage for the joint broadband PSD modelling, but they are not strictly simultaneous. The shaded Dim interval in Fig.~\ref{fig:swift_lc} denotes the \textit{Swift}/XRT segment used for the low-frequency periodogram, rather than the full temporal extent of the pre-outburst low-flux state. The Dim broadband reconstruction therefore relies on an approximate stationarity assumption. However, the 98 d offset is much shorter than the \(\simeq1048\) d \textit{Swift}/XRT Dim baseline, and both observations belong to the pre-outburst low-flux regime rather than to the 2018 changing-look outburst.

The Dim broadband PSD is constructed by combining the pre-outburst \textit{Swift}/XRT interval with the Dim \textit{XMM-Newton} observation. For the outburst phase, the \textit{Swift}/XRT up-state segment characterizes the long-term low-frequency variability of the overall bright episode, while the O1 and O2 \textit{XMM-Newton} observations provide phase-specific high-frequency constraints at the peak and decay epochs, respectively. We therefore construct two broadband PSD reconstructions by combining the same \textit{Swift}/XRT up-state segment with either the O1 or the O2 \textit{XMM-Newton} observation.

The \textit{Swift}/XRT up-state segment should not be interpreted as a purely O1 or purely O2 low-frequency PSD. Instead, it provides the only available long-baseline constraint on the outburst-timescale low-frequency variability. The O1 and O2 broadband reconstructions are therefore used to test whether the phase-specific \textit{XMM-Newton} high-frequency variability at the peak and decay epochs can be connected to this common outburst-timescale variability within a single stationary PSD parametrization. For this reason, throughout the paper we refer to these as O1-associated and O2-associated broadband reconstructions, rather than as independent low-frequency PSD measurements of the O1 and O2 phases. The segment definitions used in the broadband reconstructions are summarized in Table~\ref{tab:obs_summary}.

\begin{table*}
\centering
\caption{Observational segments used for the broadband PSD reconstructions. The O1- and O2-associated reconstructions use the same \textit{Swift}/XRT up-state monitoring segment as the low-frequency constraint, but different \textit{XMM-Newton} observations as the high-frequency constraint.}
\label{tab:obs_summary}
\begingroup\small
\begin{tabular}{@{}lp{0.16\textwidth}lp{0.15\textwidth}p{0.31\textwidth}@{}}
\hline
Reconstruction & \textit{Swift}/XRT MJD range & \textit{XMM-Newton} ObsID & \textit{XMM} epoch & Role in the joint PSD analysis \\
\hline
Dim & 56185--57233 & 0763500201 & pre-outburst low state & independent Dim low- and high-frequency reconstruction \\
O1-associated & 58293--58437 & 0800840201 & outburst peak & common up-state low-frequency constraint plus O1 high-frequency constraint \\
O2-associated & 58293--58437 & 0820530401 & post-peak decay & common up-state low-frequency constraint plus O2 high-frequency constraint \\
\hline
\end{tabular}
\endgroup
\begin{flushleft}
\footnotesize
Notes. The \textit{Swift}/XRT sampling windows, temporal gaps, and individual observing times are applied directly in the PSRESP simulations. The up-state segment has typical visit separations of several days to about one week, while the longer Dim segment contains larger temporal gaps; individual \textit{Swift}/XRT visits have typical effective exposures of order $10^3$ s. The O1- and O2-associated rows share the same long-baseline \textit{Swift}/XRT up-state segment; they differ only in the high-frequency \textit{XMM-Newton} observation used in the joint reconstruction.
\end{flushleft}
\end{table*}

\begin{figure}
    \centering
    \includegraphics[width=\columnwidth]{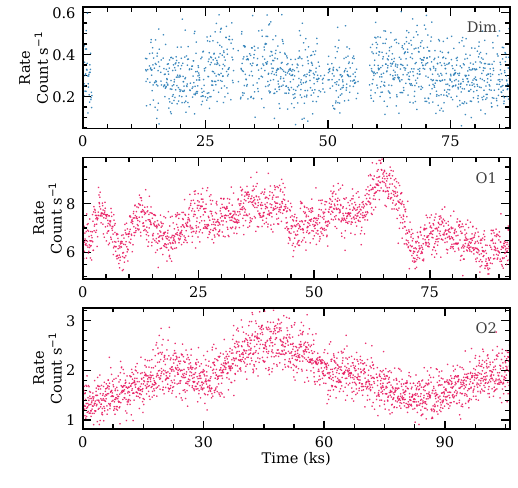}
        \caption{\textit{XMM-Newton} 2--10 keV light curve of NGC 1566.}
    \label{fig:xmm_lc}
\end{figure}

\subsection{PSD modelling and joint constraints}

We model the intrinsic PSD with a bending power law,
\begin{equation}
P(f)=A
\frac{(f/f_{\rm b})^{-\alpha_{\rm low}}}
{1+(f/f_{\rm b})^{\alpha_{\rm high}-\alpha_{\rm low}}},
\label{eq:bending_psd}
\end{equation}
where \(A\) is a normalization factor, \(\alpha_{\rm low}\) and \(\alpha_{\rm high}\) are the low- and high-frequency slopes, and \(f_{\rm b}\) is the bend frequency. The corresponding characteristic timescale is \(t_{\rm b}=1/f_{\rm b}\). In the bending-power-law grid search, \(\alpha_{\rm low}\), \(f_{\rm b}\), and \(\alpha_{\rm high}\) are all treated as free grid parameters. The bend-frequency grid was sampled uniformly in \(\log_{10}(f_{\rm b}/{\rm Hz})\), with spacings of 0.02 dex for the Dim-state models and 0.01 dex for the O1- and O2-associated models. The grid boundaries were expanded where necessary to ensure that the preferred and accepted solutions were not set by the initial parameter limits. The reported preferred \(f_{\rm b}\) values therefore correspond to discrete grid nodes. When converted to Hz, the number of significant figures reflects the selected \(\log f_{\rm b}\) grid points rather than the precision of a continuous fit.

We constrain the PSD using a PSRESP-style Monte Carlo forward-modelling procedure \citep{UttleyMcHardyPapadakis2002}. For each trial PSD model, artificial light curves are generated following \citet{TimmerKoenig1995}, with simulated durations longer than the observed segments in order to account for red-noise leakage. The simulated light curves are then passed through the same observing windows and periodogram procedures as the data. This approach propagates the effects of finite duration, irregular sampling, temporal gaps, red-noise leakage, aliasing, and measurement noise into the simulated periodogram ensemble \citep[e.g.][]{Vaughan2003}.

For the irregularly sampled \textit{Swift}/XRT monitoring data, we compute Lomb--Scargle periodograms \citep{Lomb1976,Scargle1982}. The simulated \textit{Swift}/XRT light curves are sampled at the actual \textit{Swift}/XRT observing times before the same Lomb--Scargle periodogram calculation is applied. For the \textit{XMM-Newton} observations, we compute FFT periodograms from the 50 s binned EPIC-pn light curves. The simulated \textit{XMM-Newton} light curves are generated with the same duration, time binning, and good-time windows as the data, and are analysed with the same FFT-based procedure. The observed and simulated periodograms are then binned on the same logarithmic frequency grids for each data set.

Before the periodograms are computed, the simulated light curves are rescaled separately for each data set using the observed fractional variability measure adopted for that data set. For the \textit{Swift}/XRT simulations, the rescaling is determined by the observed fractional intrinsic variance of the corresponding monitoring segment. For the \textit{XMM-Newton} simulations, the rescaling is determined by the observed fractional rms relative to the mean count rate of the corresponding observation. The observed and simulated periodograms are then compared in fractional-rms units. No additional free multiplicative normalization factor is introduced in the PSRESP distance statistic.

For each data set \(d\), where \(d\) denotes either the \textit{Swift}/XRT or the \textit{XMM-Newton} periodogram, we define the PSRESP distance statistic as

\begin{equation}
D_{\rm obs,d}=\sum_i
\frac{\left[P_{\rm obs,d}(f_i)-\langle P_{\rm sim,d}(f_i)\rangle\right]^2}
{\sigma_{\rm sim,d}^2(f_i)},
\label{eq:d_obs}
\end{equation}
where \(P_{\rm obs,d}(f_i)\) is the observed binned periodogram, and \(\langle P_{\rm sim,d}(f_i)\rangle\) and \(\sigma_{\rm sim,d}(f_i)\) are the mean and standard deviation of the simulated binned periodogram powers in the \(i\)-th frequency bin. For each Monte Carlo realization \(k\), an analogous statistic \(D^{(k)}_{\rm sim,d}\) is calculated by replacing \(P_{\rm obs,d}(f_i)\) with the simulated binned periodogram \(P^{(k)}_{\rm sim,d}(f_i)\).
The broadband constraint is obtained by requiring the same intrinsic PSD shape to describe both the low-frequency \textit{Swift}/XRT periodogram and the high-frequency \textit{XMM-Newton} periodogram. Because the simulated light curves are rescaled separately for each data set before the periodogram comparison, the joint PSRESP statistic primarily tests the broadband PSD shape after the observed sampling windows and noise treatment are applied, rather than fitting the absolute rms normalization as an additional free parameter. Equation~\ref{eq:d_obs} is invariant under a common multiplicative rescaling of both the observed and simulated periodogram powers within a given data set, because the squared difference in the numerator and \(\sigma_{\rm sim,d}^{2}\) scale as the square of that factor. This invariance, however, is not equivalent to fitting an independent normalization for each trial PSD model.
For each trial model, the observed joint distance is defined as
\begin{equation}
D_{\rm obs,joint}=D_{\rm obs,Swift}+D_{\rm obs,XMM},
\label{eq:d_obs_joint}
\end{equation}
and the corresponding Monte Carlo joint distances are
\begin{equation}
D^{(k)}_{\rm sim,joint}=D^{(k)}_{\rm sim,Swift}+D^{(k)}_{\rm sim,XMM}.
\label{eq:d_sim_joint}
\end{equation}
The joint PSRESP acceptance probability is then
\begin{equation}
P_{\rm acc,joint}=\frac{N\left(D^{(k)}_{\rm sim,joint}\ge D_{\rm obs,joint}\right)}{N_{\rm sim}},
\label{eq:pacc_joint}
\end{equation}
where \(N_{\rm sim}\) is the number of Monte Carlo realizations. Thus, \(P_{\rm acc,joint}\) measures the fraction of simulated broadband periodograms that deviate from the model at least as strongly as the observed broadband periodogram. We also compute \(P_{\rm acc,Swift}\) and \(P_{\rm acc,XMM}\) separately using the same definition for the individual data sets.

   The grid search is performed using a multi-stage rechecking procedure. We first carry out a broad search over the bending-power-law parameter grid and retain the 1000 highest-ranked candidate models for each phase and energy band. These candidates are then re-evaluated with the updated PSRESP settings using \(N_{\rm sim}=150\) Monte Carlo realizations. The 300 models with the highest joint acceptance values are subsequently rechecked using \(N_{\rm sim}=500\) realizations. Finally, the 10 highest-acceptance models from this top-300 set are verified with \(N_{\rm sim}=1000\) realizations, and the model with the highest \(P_{\rm acc,joint}\) in this final check is adopted as the preferred grid solution.
In the \textit{Swift}/XRT simulations, the synthetic light curves are sampled at the actual observing times, so that the irregular cadence and long temporal gaps are propagated into the simulated periodograms. We require 10 periodogram points per binned frequency bin, and generate simulated light curves with a duration 100 times longer than the observed baseline in order to account for red-noise leakage. The simulation time step is set by dividing the shortest relevant sampling interval by 10, with an upper cap of 1000 s. For the PSD comparison plots, the forward-modelled periodogram envelope of the preferred model is generated using \(N_{\rm sim}=2000\) simulations.
We regard models with \(P_{\rm acc,joint}\geq0.05\) as acceptable in the PSRESP sense. Models below this threshold are not treated as providing robust measurements of the bend frequency. This criterion is applied uniformly to all phases and energy bands.
For each broadband reconstruction and energy band, the preferred value reported in Table~\ref{tab:joint_psresp_bending} corresponds to the highest-acceptance solution after the final \(N_{\rm sim}=1000\) verification of the best 10 candidates. The quoted acceptance probabilities are therefore reported at the precision allowed by this final verification step; values smaller than one successful realization are listed as \(<10^{-3}\). The bracketed ranges in Table~\ref{tab:joint_psresp_bending} are derived from the accepted models in the \(N_{\rm sim}=500\) top-300 recheck. These ranges should therefore be interpreted as empirical accepted-model ranges within the PSRESP rechecking procedure, rather than as likelihood-based confidence intervals or as the result of a full posterior exploration. For comparisons between states or outburst phases, we construct empirical distributions of the relevant \(\Delta\log f_{\rm b}\) values from the accepted rechecked grid solutions, and quote the corresponding central 90 per cent empirical ranges only when both compared phases have acceptable broadband PSD solutions.

\begin{table*}
\centering
\caption{Broadband bending-power-law PSD parameters from the joint \textit{Swift}+\textit{XMM-Newton} PSRESP analysis.}
\label{tab:joint_psresp_bending}
\begin{tabular}{llcccccc}
\hline
Reconstruction & Band &
$\alpha_{\rm low}$ &
$\log(f_{\rm b}/{\rm Hz})$ &
$\alpha_{\rm high}$ &
$P_{\rm acc,Swift}$ &
$P_{\rm acc,XMM}$ &
$P_{\rm acc,joint}$ \\
\hline
Dim & Soft &
$0.65\,[0.50,0.85]$ &
$-4.710\,[-4.730,-4.430]$ &
$1.45\,[1.25,1.45]$ &
0.730 & 0.564 & 0.762 \\

O1-associated$^{\dagger}$ & Soft &
$1.05$ &
$-6.332$ &
$1.65$ &
0.450 & $<10^{-3}$ & 0.026 \\

O2-associated & Soft &
$1.05\,[0.95,1.15]$ &
$-6.562\,[-6.602,-6.302]$ &
$1.70\,[1.65,1.85]$ &
0.780 & 0.856 & 0.906 \\

Dim & Hard &
$0.80\,[0.70,0.95]$ &
$-4.670\,[-4.730,-4.430]$ &
$1.30\,[1.25,1.45]$ &
0.744 & 0.938 & 0.870 \\

O1-associated$^{\dagger}$ & Hard &
$1.00$ &
$-6.422$ &
$1.65$ &
0.612 & $<10^{-3}$ & $8\times10^{-3}$ \\

O2-associated & Hard &
$0.90\,[0.80,1.00]$ &
$-6.562\,[-6.602,-6.302]$ &
$1.80\,[1.65,1.85]$ &
0.812 & 0.822 & 0.926 \\
\hline
\end{tabular}

\begin{flushleft}
\footnotesize
Notes. The first value in each parameter column gives the preferred grid solution, corresponding to the highest-acceptance model after the final \(N_{\rm sim}=1000\) verification of the best 10 candidates. Bracketed intervals give the parameter span of the accepted models with \(P_{\rm acc,joint}\geq0.05\) in the \(N_{\rm sim}=500\) top-300 recheck, and should be interpreted as empirical accepted-model ranges rather than formal confidence intervals. For the O1 peak observation, no bending-power-law solution satisfies this adopted joint acceptance criterion; the daggered entries are therefore shown only as diagnostic highest-acceptance grid solutions and are not used as robust bend-frequency measurements. For the O2-associated reconstruction, the identical preferred soft- and hard-band bend frequencies correspond to the same discrete grid point, \(\log_{10}(f_{\rm b}/{\rm Hz})=-6.562\), in the bend-frequency search. The same accepted range, \([-6.602,-6.302]\), likewise reflects the rechecked acceptable grid nodes. This coincidence should not be interpreted as evidence for exactly identical physical bend frequencies in the two energy bands, but rather as the absence of a resolved soft--hard difference within the adopted grid resolution and frequency coverage.
\end{flushleft}
\end{table*}

\section{Results}

\subsection{Broadband PSRESP constraints}

Table~\ref{tab:joint_psresp_bending} summarizes the joint \textit{Swift}+\textit{XMM-Newton} PSRESP constraints obtained with the bending-power-law PSD model. The three broadband reconstructions show different levels of consistency within this framework. The Dim reconstruction and the O2-associated reconstruction can be described by acceptable broadband bending-power-law solutions in both the soft and hard bands, whereas the O1-associated reconstruction is not acceptably represented by a single stationary broadband bending-power-law PSD.

Because the same \textit{Swift}/XRT up-state segment is used for the O1- and O2-associated reconstructions, the comparison between them should not be interpreted as a measurement of two independent outburst low-frequency PSDs. Rather, it tests whether each phase-specific \textit{XMM-Newton} high-frequency observation can be connected to the common outburst-timescale low-frequency variability within the same stationary PSD parametrization.

This distinction is clearest in the joint acceptance statistic. For the O1 peak observation, the highest-acceptance grid solutions have \(P_{\rm acc,joint}=0.026\) in the soft band and \(8\times10^{-3}\) in the hard band. Both values fall below the adopted acceptance threshold of \(P_{\rm acc,joint}=0.05\). This indicates that the \textit{Swift}/XRT long-timescale periodogram and the O1 \textit{XMM-Newton} high-frequency periodogram cannot be simultaneously reproduced by a single stationary broadband bending-power-law PSD. We therefore list the O1-associated parameters in Table~\ref{tab:joint_psresp_bending} only as diagnostic highest-acceptance grid solutions, and do not use them as robust measurements of \(f_{\rm b}\).

\begin{figure*}
    \centering
    \includegraphics[width=\textwidth]{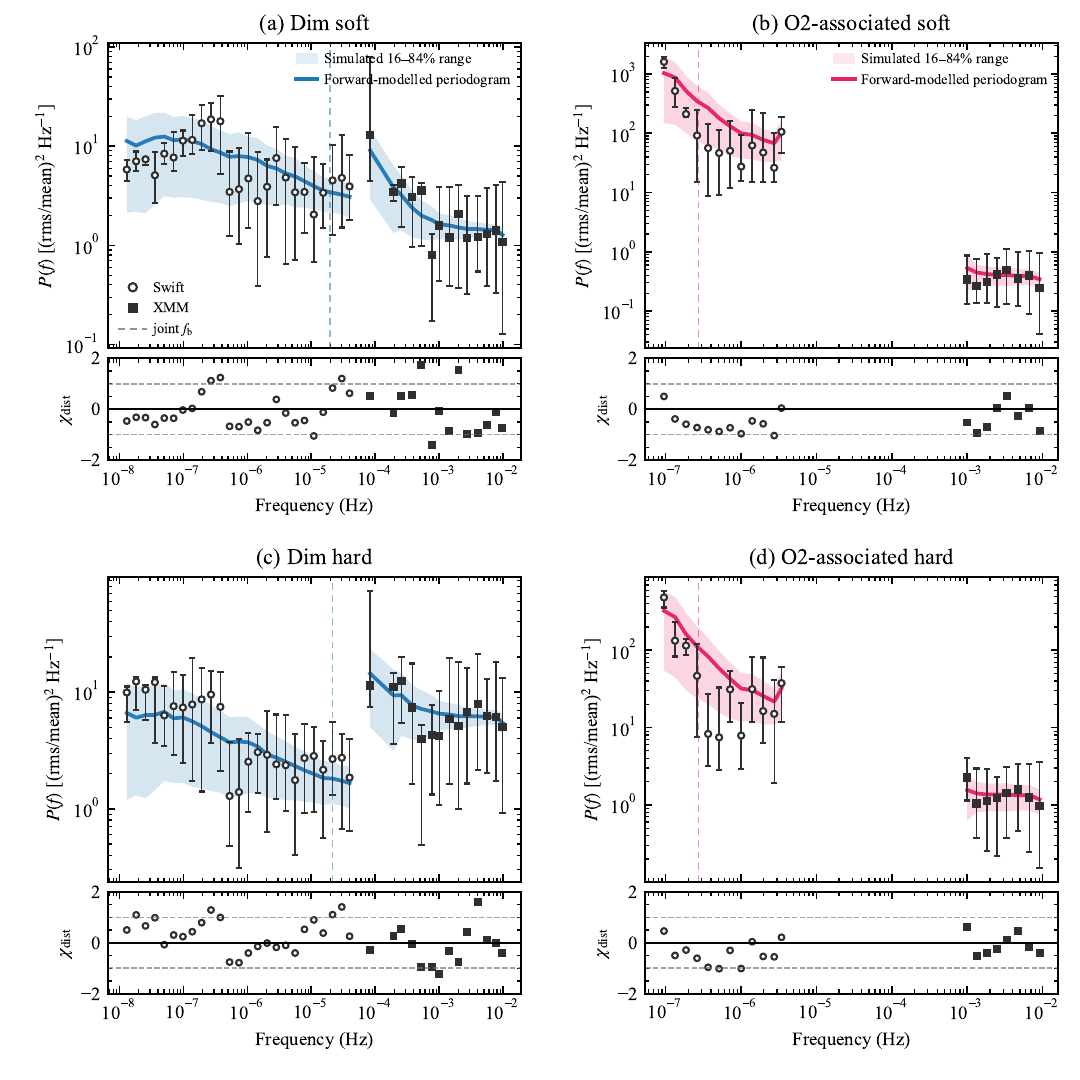}
    \caption{
    Broadband periodogram comparison for the accepted Dim and O2-associated bending-power-law solutions. 
    Open circles and filled squares show the binned \textit{Swift}/XRT and \textit{XMM-Newton} periodograms, respectively. 
    Solid curves and shaded regions show the median and 16th--84th percentile range of the forward-modelled simulated periodograms generated from the highest-acceptance PSRESP solution after applying the same sampling windows, noise treatment, and periodogram procedure as used for the data. 
    Vertical dashed lines mark the joint \textit{Swift}+\textit{XMM-Newton} bend frequencies listed in Table~\ref{tab:joint_psresp_bending}. 
    The lower panels show the signed residuals, defined as \(\chi_{{\rm dist},i}=[P_{\rm obs}(f_i)-\langle P_{\rm sim}(f_i)\rangle]/\sigma_{\rm sim}(f_i)\).
    }
    \label{fig:periodogram_psresp}
\end{figure*}

Figure~\ref{fig:periodogram_psresp} shows the periodogram-level comparison for the accepted Dim and O2-associated solutions. The binned \textit{Swift}/XRT and \textit{XMM-Newton} periodograms are generally consistent with the forward-modelled periodogram distributions from the highest-acceptance PSRESP solutions. For the O2-associated reconstruction, the binned \textit{Swift}/XRT periodograms cover approximately \(9.5\times10^{-8}\)--\(3.4\times10^{-6}\,\mathrm{Hz}\) in both the soft and hard bands, whereas the \textit{XMM-Newton} periodograms cover approximately \(1.0\times10^{-3}\)--\(9.1\times10^{-3}\,\mathrm{Hz}\). The unsampled interval between the two frequency windows therefore spans a factor of \(\sim290\), or about \(2.5\) dex. The preferred O2-associated bend frequency, \(\log_{10}(f_{\rm b}/{\rm Hz})=-6.562\), corresponding to \(f_{\rm b}=2.74\times10^{-7}\,\mathrm{Hz}\), lies within the low-frequency \textit{Swift}/XRT window rather than inside the \textit{Swift}--\textit{XMM-Newton} gap. However, it lies a factor of \(\sim3.6\times10^{3}\) below the lower edge of the \textit{XMM-Newton} frequency range. Thus, the \textit{XMM-Newton} data mainly constrain the high-frequency PSD behaviour within the adopted bending-power-law model, while the bend location is constrained primarily by the \textit{Swift}/XRT window and by the joint broadband PSD reconstruction. The vertical dashed lines mark the joint \textit{Swift}+\textit{XMM-Newton} bend frequencies listed in Table~\ref{tab:joint_psresp_bending}.

\subsection{Bend-frequency evolution from the Dim phase to the O2-associated reconstruction}

The accepted Dim and O2-associated solutions occupy distinct bend-frequency ranges. In the soft band, the highest-acceptance solution shifts from \(\log(f_{\rm b}/{\rm Hz})=-4.710\) in the Dim phase to \(-6.562\) in the O2-associated reconstruction, corresponding to \(f_{\rm b}=1.95\times10^{-5}\) and \(2.74\times10^{-7}\,\mathrm{Hz}\), respectively. In the hard band, the corresponding shift is from \(\log(f_{\rm b}/{\rm Hz})=-4.670\) to \(-6.562\), or from \(2.14\times10^{-5}\) to \(2.74\times10^{-7}\,\mathrm{Hz}\). The full accepted parameter ranges and acceptance values are given in Table~\ref{tab:joint_psresp_bending}.

Using the highest-acceptance grid solutions, the Dim-to-O2-associated bend-frequency displacement is
\begin{equation}
\Delta\log f_{\rm b}=\log f_{\rm b,O2}-\log f_{\rm b,Dim}=-1.85
\end{equation}
in the soft band and \(-1.89\) in the hard band. The accepted-model combinations from the Dim and O2-associated accepted grids give central 90 per cent empirical ranges of \(\Delta\log f_{\rm b}=[-2.08,-1.65]\) in both bands. Thus, within the accepted PSRESP region, the O2-associated broadband reconstruction is consistently shifted to lower characteristic frequencies than the Dim reconstruction.

In terms of the characteristic timescale \(t_{\rm b}=1/f_{\rm b}\), the highest-acceptance solutions correspond to \(t_{\rm b}=0.59\) d in the Dim soft band, \(0.54\) d in the Dim hard band, and \(42.2\) d in both O2-associated bands. The implied timescale increase is therefore a factor of \(\simeq70\)--80. This displacement appears in both energy bands, suggesting that the inferred timing change is not confined to a single spectral component.

\subsection{Energy dependence and high-frequency slopes}

The soft and hard bands give closely consistent Dim-to-O2-associated bend-frequency shifts. The dominant state dependence is therefore a broadband shift of the characteristic frequency, rather than a strong change in the relative soft--hard bend-frequency separation. This is different from a scenario in which the timing evolution is driven mainly by one energy band.

The preferred O2-associated values of \(\alpha_{\rm high}\) are numerically larger than the corresponding Dim values in both energy bands. However, we do not treat this numerical difference as an independent physical result. Within the accepted grid region, \(\alpha_{\rm high}\) and \(f_{\rm b}\) are partially degenerate: a change in the bend location can alter the frequency range over which the high-frequency slope is effectively constrained by the \textit{XMM-Newton} periodogram. We therefore interpret the high-frequency slopes only as part of the phenomenological PSD description, rather than as an independent geometric measurement. The more robust inference is that the accepted bend-frequency region moves systematically to lower frequencies in both energy bands in the O2-associated reconstruction.

\begin{figure*}
    \centering
    \includegraphics[width=\textwidth]{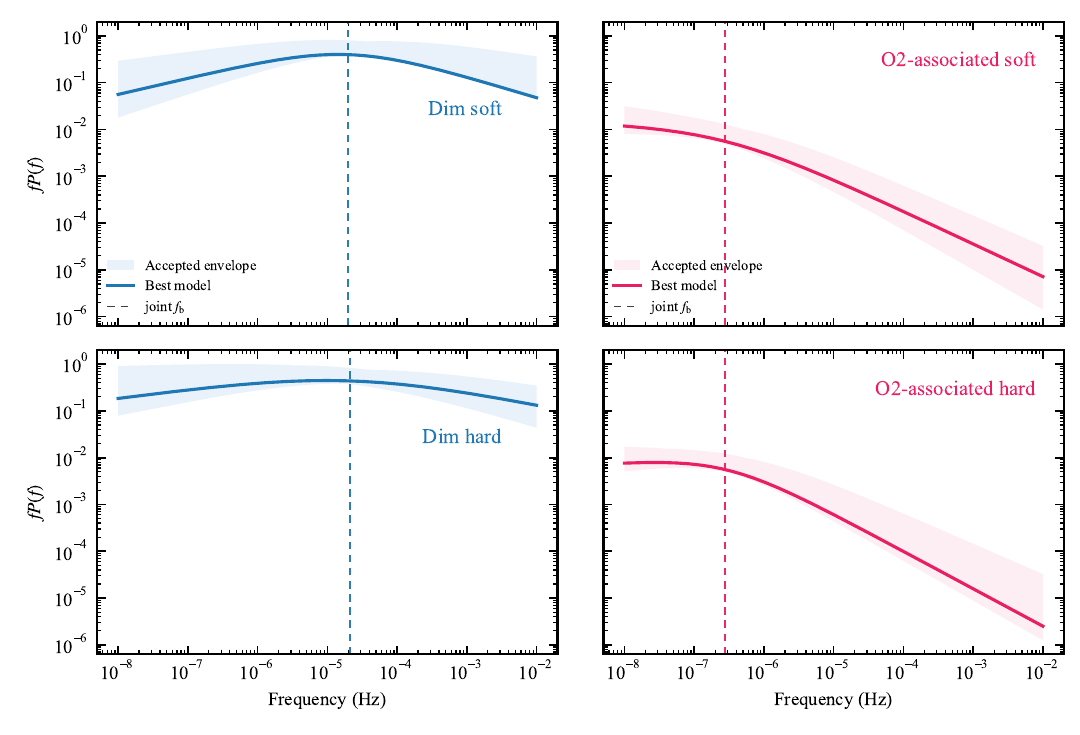}
    \caption{
    Underlying bending-power-law PSD models for the accepted Dim and O2-associated solutions, shown in \(fP(f)\) space. 
    Solid curves show the highest-acceptance models, shaded regions show the range spanned by the accepted PSRESP grid solutions, and vertical dashed lines mark the joint bend frequencies in Table~\ref{tab:joint_psresp_bending}. 
    The O2-associated models are shifted to lower characteristic frequencies than the Dim models in both energy bands within the adopted bending-power-law framework.
    }
    \label{fig:underlying_bending_psd}
\end{figure*}

Figure~\ref{fig:underlying_bending_psd} shows the corresponding underlying bending-power-law PSD models in \(fP(f)\) space. This representation highlights the displacement of the accepted model envelopes. The Dim models peak at higher characteristic frequencies, whereas the O2-associated models are shifted to lower frequencies in both the soft and hard bands.

\subsection{Supplementary soft--hard timing check}
As a supplementary check on the relationship between the soft and hard bands, we computed the soft--hard cross spectrum from the 50 s binned \textit{XMM-Newton} light curves. We define the quoted lag as
\begin{equation}
\tau_{\rm s-h}=t_{\rm soft}-t_{\rm hard},
\end{equation}
so that a negative value corresponds to the soft-band variations leading the hard-band variations under this convention. Because each phase is represented by a single observation, the number of independent 20 ks segments is limited: one for the Dim observation, three for O1 peak, and two for O2 decay. The resulting coherence and lag estimates should therefore be regarded as diagnostic checks rather than as independent quantitative constraints on the PSD modelling.

At the lowest Fourier frequencies, where the estimates are most stable, the O1 peak and O2 decay observations show high soft--hard coherence. Averaged over \(5.0\times10^{-5}\)--\(2.25\times10^{-4}\,\mathrm{Hz}\), the mean coherence is \(\simeq0.91\) for O1 peak and \(\simeq0.87\) for O2 decay. The Dim observation is less well constrained because only one 20 ks segment is available; over the same frequency range, the mean coherence is \(\simeq0.23\), with large uncertainty expected from the small number of independent segments.

Using the same low-frequency range, the inverse-variance weighted soft-minus-hard lags are $659\pm799$ s for the Dim observation, $-234\pm50$ s for the O1 peak, and $-103\pm82$ s for the O2 decay observation. These uncertainties are formal statistical errors from the averaged cross spectra. For cross-spectral measurements based on $N$ independent segments, the phase and lag uncertainties depend on both $N$ and the coherence, with the phase uncertainty approximately scaling as $\sigma_\phi \propto [(1-\gamma^2)/(2\gamma^2 N)]^{1/2}$, where $\gamma^2$ is the coherence. Because the O1 and O2 estimates are based on only three and two independent 20 ks segments, respectively, the formal errors likely underestimate the true uncertainty, which includes contributions from realization scatter, red-noise leakage, covariance between neighbouring frequency bins, and possible non-stationarity of the underlying variability process.

As a robustness check, we repeated the O1 cross-spectral analysis using individual 20 ks segments, leave-one-out combinations, and alternative segment lengths. The low-frequency lag remains negative in these tests, but its amplitude varies substantially. This behaviour indicates that the sign of the O1 lag is relatively stable in the adopted low-frequency range, but that its amplitude is not robustly determined. The O2 lag amplitude should be regarded as similarly uncertain, and even less well constrained in terms of independent segment number, because it is based on only two independent 20 ks segments. Under our convention, the negative O1 and O2 lags correspond to the soft band leading the hard band. Given the small number of independent segments, this sign should not be assigned a unique physical interpretation. In particular, because the O1-associated broadband PSD is not acceptably described by a single stationary bending-power-law reconstruction, any cross-spectral lag derived from this observation may also be affected by non-stationarity or changing contributions from multiple variable components. We therefore treat the cross-spectral lags only as supplementary diagnostics of soft--hard coupling, and instead base the main physical interpretation on the PSRESP broadband PSD comparison. Although such soft-leading lags are opposite to the hard-lag behaviour expected in the simplest inward propagation-fluctuation picture, the small number of independent segments and possible non-stationarity prevent us from assigning a unique physical interpretation to the measured lag sign.

\section{Discussion and conclusions}

The main timing result of this work is that, within the bending-power-law parametrization, the accepted Dim and O2-associated broadband reconstructions show a common shift of the characteristic bend frequency to lower values in both the soft and hard X-ray bands. The preferred bend frequency changes from \(1.95\times10^{-5}\) to \(2.74\times10^{-7}\,\mathrm{Hz}\) in the soft band and from \(2.14\times10^{-5}\) to \(2.74\times10^{-7}\,\mathrm{Hz}\) in the hard band. The corresponding Dim-to-O2-associated shifts are \(\Delta\log f_{\rm b}=-1.85\) and \(-1.89\), with central 90 per cent empirical accepted-model ranges of \([-2.08,-1.65]\) in both bands. Thus, within the accepted PSRESP region, the O2-associated reconstruction favours a substantially longer characteristic X-ray variability timescale than the Dim reconstruction.

The bending-power-law PSD used here should be regarded as a phenomenological parametrization of the broadband PSD shape, rather than as a unique physical model. We also tested single-power-law PSD models as simpler alternatives. These tests, summarized in Appendix~\ref{app:single_power_law}, show that single-power-law models can yield non-negligible PSRESP acceptance values in several cases. The present data therefore do not formally exclude a single-power-law description for every individual reconstruction. However, a single-power-law model does not provide a characteristic bend frequency, and hence cannot quantify the state-dependent shift in \(f_{\rm b}\) that is the focus of this work. We therefore use the bending-power-law model as a common phenomenological framework for comparing characteristic frequencies between the Dim and O2-associated reconstructions.

In this sense, the inferred \(f_{\rm b}\) should be interpreted as the characteristic frequency within the adopted bending-power-law framework, not as a model-independent measurement of a unique physical PSD break. More complex models, such as sharply broken power laws, double-bend PSDs, or Lorentzian components \citep[e.g.][]{Belloni2010}, may be relevant for denser and more continuous data sets. In the present case, however, the available \textit{Swift}+\textit{XMM-Newton} data mainly constrain separated frequency ranges and do not provide enough leverage to determine additional shape parameters robustly. The bending-power-law model is therefore used as a controlled baseline for describing the Dim and O2-associated broadband reconstructions, while the single-power-law tests are treated as a check on model dependence.

The lower preferred $f_{\rm b}$ in the O2-associated reconstruction appears counterintuitive when compared with the empirical McHardy et al. (2006) scaling, which predicts shorter characteristic timescales, or equivalently higher bend frequencies, at higher bolometric luminosity for a fixed black-hole mass. For NGC 1566, published black-hole mass estimates are of order a few \(10^{6}\,M_{\odot}\), with representative values including \(8.3\times10^{6}\,M_{\odot}\) \citep{Woo2002}, \((8.6\pm4.4)\times10^{6}\,M_{\odot}\) \citep{Smajic2015}, and a compiled mean value of \((5.3\pm2.7)\times10^{6}\,M_{\odot}\) \citep{Ochmann2024}. These estimates indicate an uncertainty of approximately a factor of a few in the black-hole mass of NGC 1566. This mass-related uncertainty is nevertheless smaller than the observed timing displacement between the Dim and O2-associated reconstructions. Using the McHardy relation as a simple benchmark, \(T_{\rm B}\propto M_{\rm BH}^{2.1}\). Therefore, a factor-of-two uncertainty in \(M_{\rm BH}\) would propagate to an uncertainty of \(2.1\log_{10}2\simeq0.63\) dex in the predicted timescale, or equivalently in the opposite direction in the predicted bend frequency. By comparison, the preferred bend frequency shifts by \(\simeq1.85\) dex in the soft band and \(\simeq1.89\) dex in the hard band between the Dim and O2-associated reconstructions. Thus, the observed shift is substantially larger than can be produced by the black-hole mass uncertainty alone, and is instead consistent with a genuine state-dependent change in the characteristic X-ray variability timescale within the adopted bending-power-law framework.

Given these uncertainties, and because both the Dim and O2-associated reconstructions correspond to selected states of a recurrent changing-look AGN, the McHardy scaling should be regarded only as a heuristic benchmark rather than a quantitative test. In particular, the outburst reconstructions combine a common Swift/XRT up-state monitoring segment with phase-specific XMM-Newton observations, and the assumptions underlying a quasi-stationary Seyfert-like timing scaling may not apply cleanly to either endpoint of the Dim--O2 comparison.

Under this interpretation, the observed bend-frequency displacement primarily reflects a state-dependent reconfiguration of the broadband X-ray variability structure, rather than a straightforward one-parameter accretion-rate effect.

Dedicated broadband spectral studies have shown that the 2018 outburst involved substantial changes in the inner radiative structure. \citet{Jana2021} found a strong soft X-ray excess during the outburst and interpreted it as emission from a warm Comptonizing region associated with the inner disc. \citet{TripathiDewangan2022} further showed that, during the outburst decline, the seed-photon supply to the hot corona and the coronal properties changed strongly: the hot-corona electron temperature increased from \(\sim22\) to \(\sim200\) keV, the optical depth decreased from \(\tau_{\rm hot}\sim4\) to \(\sim0.7\), and the scattering fraction increased from \(\sim1\) per cent to \(\sim10\) per cent. These spectral changes indicate that the disc, soft-excess component, and hot corona evolved together during the decay stage.

In this context, the lower \(f_{\rm b}\) in the O2-associated reconstruction is more naturally understood as a timing signature of disc--corona reconfiguration than as a direct consequence of a single global accretion-rate parameter. A stronger or more extended Comptonizing medium, a changing seed-photon cooling rate, enhanced scattering or reprocessing, or a larger effective region contributing to the X-ray variability could smooth short-timescale fluctuations and move the preferred characteristic frequency to lower values. The fact that the shift is seen in both the soft and hard bands supports this interpretation: the effect is not confined to the soft-excess component, but affects the broadband X-ray variability structure.

The O1 peak observation provides a useful diagnostic, but no robust bend-frequency measurement can be obtained from this epoch. In the bending-power-law analysis, the O1-associated reconstruction does not satisfy the adopted joint PSRESP acceptance threshold in either energy band. Consequently, the O1 XMM-Newton high-frequency periodogram cannot be connected to the common outburst-timescale Swift/XRT variability by a single stationary broadband bending-power-law PSD with a well-constrained characteristic bend.

A possible explanation for the lack of a robust O1 bend-frequency
measurement is that the relevant characteristic frequency is poorly
sampled by the available frequency windows. To quantify this point,
we used the empirical scaling relation of \citet{McHardy2006},
\begin{equation}
\log T_{\rm B}
=
2.10 \log \left( \frac{M_{\rm BH}}{10^{6} M_{\odot}} \right)
-
0.98 \log \left( \frac{L_{\rm bol}}{10^{44}~{\rm erg~s^{-1}}} \right)
-
2.32 ,
\end{equation}
where \(T_{\rm B}\) is in days. For the O1 peak epoch,
\citet{Jana2021} reported \(L_{\rm bol}=7.11\times10^{43}~{\rm erg~s^{-1}}\)
and adopted \(M_{\rm BH}=8.3\times10^{6}M_{\odot}\). These values give
\(T_{\rm B}\simeq0.57~{\rm d}\), corresponding to
\(f_{\rm b}\simeq2.0\times10^{-5}~{\rm Hz}\). The peak luminosity of
approximately \(5\) per cent of \(L_{\rm Edd}\) estimated by
\citet{Parker2019} gives a similar estimate: for
\(M_{\rm BH}=10^{7}M_{\odot}\), it corresponds to
\(L_{\rm bol}\simeq6.3\times10^{43}~{\rm erg~s^{-1}}\) and predicts
\(f_{\rm b}\simeq1.2\times10^{-5}~{\rm Hz}\).

These characteristic frequencies lie above the highest effective
Swift/XRT frequencies of the up-state monitoring segment,
\(\simeq3.4\times10^{-6}~{\rm Hz}\), but below the lower edge of the
XMM-Newton periodogram window, \(\simeq10^{-3}~{\rm Hz}\). The O1
bend expected from this heuristic scaling would therefore fall in the
unsampled Swift--XMM frequency gap. Equivalently, since
\(f_{\rm b}\propto T_{\rm B}^{-1}\), the McHardy scaling implies
\(f_{\rm b}\propto L_{\rm bol}^{0.98}M_{\rm BH}^{-2.10}\). A factor of
two uncertainty in \(L_{\rm bol}\) changes the predicted frequency by
\(\simeq0.3\) dex, whereas a factor of two uncertainty in \(M_{\rm BH}\)
changes it by \(\simeq0.6\) dex. Thus, bolometric-correction
uncertainties, black-hole-mass uncertainty, and the intrinsic scatter
of the McHardy relation affect the numerical prediction, but not the
basic conclusion that the expected O1 bend is not directly sampled by
the available frequency windows. This provides a plausible explanation,
in terms of frequency coverage, for why the O1-associated reconstruction
does not yield a robust bend-frequency measurement, especially given
the additional complications of non-stationarity during the peak
outburst phase and possible multi-component variability.

This interpretation is consistent with the single-power-law checks in Appendix A, which yield non-negligible joint PSRESP acceptance values for the O1-associated reconstruction, including an acceptable hard-band value of $P_{\rm acc,joint}=0.095$. Therefore, the O1-associated result reflects the absence of a robust bend-frequency measurement within the current frequency coverage, rather than the absence of a physical bend. Possible explanations include an unresolved high-frequency bend, non-stationarity during the peak outburst stage, multiple variable components, or a combination of these effects.

The O1 result remains physically informative. It indicates that the peak outburst stage is not described by the same simple, well-constrained characteristic-frequency reconstruction that applies to the accepted Dim and O2-associated cases. The present data do not allow unique discrimination among an unresolved high-frequency bend, true non-stationarity, and multiple variable components.

The single-power-law checks clarify the model dependence of this interpretation. They show that the current data do not uniquely require a bending form in every individual reconstruction, especially for the O1-associated case. However, a single-power-law model does not provide a bend frequency and therefore cannot quantify a characteristic timescale or its state-dependent displacement. The single-power-law results are therefore best viewed as a caution against over-interpreting the absolute physical meaning of any individual \(f_{\rm b}\), rather than as an alternative measurement of the Dim-to-O2-associated timing shift. Within the common bending-power-law framework, the accepted Dim and O2-associated solutions still occupy clearly separated bend-frequency regions in both energy bands.

Several limitations should be kept in mind. First, each phase is represented at high frequencies by a single \textit{XMM-Newton} observation. The inferred PSD differences may therefore include contributions from stochastic realization scatter as well as from true state-dependent variability changes. The PSRESP simulations account for the observed sampling, red-noise leakage, aliasing, and measurement noise, but they cannot replace multiple independent high-time-resolution observations within the same phase. Second, the O1- and O2-associated reconstructions share the same \textit{Swift}/XRT up-state monitoring segment at low frequencies. This segment spans MJD 58293--58437, or about 144 d, and covers both the O1 peak and the O2 decay stage. During this interval the source underwent large-amplitude flux evolution and a substantial post-peak decline, so the \textit{Swift}/XRT low-frequency periodogram should be regarded as an effective description of the outburst-timescale variability rather than as a strictly stationary PSD measurement. This secular flux evolution can affect the inferred low-frequency power and may influence the preferred bend location, particularly for the O2-associated reconstruction. The O2-associated bend-frequency shift should therefore be interpreted as the result of connecting the O2 high-frequency observation to this common outburst-timescale low-frequency variability under the adopted stationary-PSD approximation, not as a completely independent O2-only low-frequency PSD measurement. Third, the \textit{Swift} and \textit{XMM-Newton} data probe separated frequency ranges, leaving a gap between the long-term monitoring and the short continuous observations. This limits the ability to distinguish a smooth bend from a single power law, a sharper break, or more complex PSD curvature. Fourth, the present work does not perform a new full broadband spectral decomposition. Instead, we interpret the timing results in the context of existing dedicated spectral studies of the 2018 outburst.

In summary, we have combined long-term \textit{Swift}/XRT monitoring with high-time-resolution \textit{XMM-Newton} observations to investigate the broadband X-ray PSD of the changing-look AGN NGC~1566. Motivated by the non-stationary nature of the 2018 outburst, we treated the O1 peak and O2 decay \textit{XMM-Newton} observations separately, rather than assuming a single stationary outburst state. The outburst-timescale \textit{Swift}/XRT monitoring segment was used as a common low-frequency constraint for the O1- and O2-associated reconstructions.

Within the PSRESP bending-power-law framework, the Dim reconstruction and the O2-associated reconstruction are acceptably described by broadband PSD solutions in both the soft and hard bands. The O2-associated reconstruction favours a systematically lower characteristic bend frequency than the Dim reconstruction, with the preferred value shifting from \(1.95\times10^{-5}\) to \(2.74\times10^{-7}\,\mathrm{Hz}\) in the soft band and from \(2.14\times10^{-5}\) to \(2.74\times10^{-7}\,\mathrm{Hz}\) in the hard band. This corresponds to a characteristic variability timescale longer by a factor of about 70--80.

The common bend-frequency displacement in both energy bands indicates that the timing evolution is not confined to the soft-excess component alone, but reflects a broader change in the X-ray variability structure. In contrast, the O1 peak observation cannot be assigned a robust bend frequency within the same single stationary broadband bending-power-law description. We therefore interpret the O1 result not as a precise timing measurement, but as evidence that the peak outburst stage is not captured by the same simple characteristic-frequency description that works for the accepted Dim and O2-associated reconstructions.

These results revise the simpler dim-versus-bright interpretation by showing that the 2018 changing-look event in NGC~1566 is not merely a luminosity rescaling or a one-parameter accretion-rate change. Instead, together with previous spectral studies, the timing results suggest a transient reconfiguration of the disc--corona variability structure, with the post-peak decay observation associated with a longer dominant broadband X-ray variability timescale. This interpretation remains limited by the sparse long-term sampling, by the common use of the \textit{Swift}/XRT up-state segment for the outburst reconstructions, and by the availability of only one high-time-resolution \textit{XMM-Newton} observation per phase. Future dense monitoring will be required to determine whether the inferred bend-frequency displacement is a persistent state property or is partly affected by stochastic realization scatter.

\section*{Acknowledgements}

This work has made use of public data from the \textit{Swift} and \textit{XMM-Newton} archives. The data reduction and timing analysis made use of SAS and standard Python scientific packages, including \textsc{NumPy}, \textsc{Pandas}, \textsc{Matplotlib}, and \textsc{Astropy}.

\section*{DATA AVAILABILITY}

The data underlying this article are publicly available from the Swift and XMM-Newton data archives. The XMM-Newton observations used in this work are available from the XMM-Newton Science Archive under ObsIDs 0763500201, 0800840201, and 0820530401. The Swift/XRT monitoring data are publicly available through the UK Swift Science Data Centre and/or the HEASARC archive.
\bibliographystyle{mnras}
\bibliography{example}
\clearpage
\appendix
\section{Single-power-law checks}
\label{app:single_power_law}

As a check on the model dependence of the PSD interpretation, we also fitted single-power-law PSD models using the same \textit{Swift}+\textit{XMM-Newton} PSRESP framework. The results are summarized in Table~\ref{tab:single_power_law}. The single-power-law models yield non-negligible empirical acceptance values in several cases, indicating that the present data do not formally require a bend in every individual reconstruction. This is particularly important for the O1-associated case: although the bending-power-law model does not provide an acceptable robust bend-frequency measurement for O1, the single-power-law checks show that this should not be over-interpreted as proof of non-stationarity. Rather, the data do not support a robust characteristic bend frequency for this epoch within the adopted bending-power-law framework.

Because a single-power-law PSD does not provide a characteristic bend frequency, these fits cannot be used to quantify the state-dependent shift in \(f_{\rm b}\). We therefore use them only as model-dependence checks, while adopting the bending-power-law model in the main analysis as a common phenomenological parametrization for comparing characteristic frequencies between the Dim and O2-associated reconstructions.

\par\medskip
\refstepcounter{table}\label{tab:single_power_law}
\noindent{\bf Table~\thetable.} Single-power-law PSD checks using the joint \textit{Swift}+\textit{XMM-Newton} PSRESP framework.
\par\smallskip
\begin{center}
\begingroup\scriptsize\setlength{\tabcolsep}{2.5pt}
\begin{tabular}{llcccc}
\hline
Reconstruction & Band & \(\alpha_{\rm PL}\) & \(P_{\rm acc,Swift}\) & \(P_{\rm acc,XMM}\) & \(P_{\rm acc,joint}\) \\
\hline
Dim & Soft & 0.80 & 0.425 & 0.187 & 0.372 \\
O1-associated & Soft & 1.20 & 0.685 & 0.003 & 0.111 \\
O2-associated & Soft & 1.30 & 0.504 & 0.425 & 0.574 \\
Dim & Hard & 0.90 & 0.353 & 0.917 & 0.444 \\
O1-associated & Hard & 1.15 & 0.602 & \(<10^{-3}\) & 0.095 \\
O2-associated & Hard & 1.15 & 0.602 & 0.382 & 0.624 \\
\hline
\end{tabular}
\endgroup
\end{center}
\noindent{\footnotesize Notes. \(\alpha_{\rm PL}\) is the highest-acceptance single-power-law slope in the PSRESP grid. The acceptance values are empirical PSRESP statistics computed in the same way as in the main bending-power-law analysis. These fits are used only as model-dependence checks; they do not provide a bend frequency and therefore do not define the characteristic timescale used in the main analysis.}


\bsp	
\label{lastpage}
\end{document}